\documentclass[journal=acsnano,manuscript=article]{achemso}

\usepackage{rotating}
\usepackage{chemformula} 
\usepackage[T1]{fontenc} 
\usepackage{amsmath}
\usepackage{graphicx}
\usepackage{bm}
\usepackage{graphicx}%
\usepackage[utf8]{inputenc}
\PassOptionsToPackage{hyphens}{url}\usepackage{hyperref}

\usepackage[nobiblatex]{xurl}
\usepackage{url}

\author{Monirul Shaikh}
\author{Aleksander L. Wysocki}
\email{wysockia@unk.edu}
\affiliation{Department of Physics and Astronomy, University of Nebraska at Kearney, Kearney, NE 68849, United States}

\title[An \textsf{achemso} demo] {Electric Polarization-Driven Modulation of Fe Adatoms on Ferroelectric $\alpha$-In$_2$Se$_3$}


\begin{document}


\begin{abstract}
The interplay among structural, electronic, and magnetic properties of Fe adatoms on the surface of two-dimensional ferroelectric $\alpha$-In$_2$Se$_3$ is investigated using first-principles electronic structure calculations, with a focus on how these properties are modulated by the direction of the electric polarization of the substrate. We identify two competing adsorption sites for Fe adatoms, whose relative stability depends on the adatom concentration and can be reversed by switching the electric polarization of $\alpha$-In$_2$Se$_3$. The calculated energy barrier for thermally activated hopping between these sites is approximately 0.4 eV, corresponding to a blocking temperature of around 100 K. The hybridization between Fe and In$_2$Se$_3$ orbitals strongly depends on the adsorption site and polarization direction, driven by variations in the local adatom geometry. As a result, the adatom’s electronic configuration, magnetic moment, and magnetic anisotropy exhibit a pronounced site dependence and can be effectively modulated by switching the electric polarization of the In$_2$Se$_3$ layer. In particular, at higher adatom concentrations, an exceptionally large perpendicular magnetic anisotropy, exceeding 200 meV per Fe atom, emerges for one polarization direction, but is largely diminished when the polarization is reversed. These findings indicate that ferroelectric substrates offer a promising route for voltage-controlled tuning of magnetic adatom properties via reversible polarization switching. 
\end{abstract}

Magnetic adatoms on surfaces~\cite{Donati2021} offer a promising platform for meeting the growing technological demand for information storage and processing at increasingly smaller scales~\cite{Natterer2017,Yang2019,Chen2023}. As the smallest possible magnetic units, these atomic-scale systems are of particular interest for ultrahigh-density magnetic memory applications~\cite{Khajetoorians2016}. In addition, magnetic adatoms are attracting significant attention as potential spin-qubit candidates for quantum information technologies~\cite{Wang2023,Phark2023,Reale2023,Reale2024,Choi2022}.

Significant experimental progress has been made in recent years, with both transition metal and rare earth adatoms being successfully deposited and characterized on a diverse range of substrates. These include metallic surfaces (e.g., Pt, Cu, and Ag)~\cite{Gambardella2003, Hirjibehedin2007, Schuh2012, Donati2014, Singha2017}; non-conductive CuN overlayer~\cite{Hirjibehedin2007}; insulating materials like MgO~\cite{Baumann2015, Donati2016, Natterer2018, Singha2021, Singha2021b} and BaO~\cite{Sorokin2023}; two-dimensional (2D) semiconductor like MoS$_2$ monolayer~\cite{Trishin2021}; graphene~\cite{Brar2011, Gyamfi2012, Eelbo2013, Baltic2016, Baltic2018, Curcella2023}; and even the 2D electron gas formed on the SrTiO$_3$ surface~\cite{Bellini2022}. In particular, Ho adatom on MgO was shown to have a significant remanent magnetization and coercivity at temperatures as high as several tens of kelvin~\cite{Donati2016}. In addition, the development of electron spin resonance scanning tunneling microscopy (ESR-STM)~\cite{Baumann2015,Seifert2021,Drost2022} has enabled the coherent control of the magnetic states of a single magnetic adatom using electromagnetic radiation, which is a key advancement toward the use of adatoms in quantum information processing~\cite{Wang2023,Phark2023,Reale2023,Reale2024,Choi2022}.

On the theoretical side, there is a substantial body of \emph{ab initio} work on magnetic adatoms on surfaces, primarily based on density functional theory (DFT). For transition metal adatoms, the DFT+U method~\cite{Anisimov1997} provides a reliable description of their electronic structure (e.g., Ref.\cite{Shehada2024}). The DFT+$U$ method has also been applied to study rare-earth adatoms (e.g.,\cite{Carbone2023}). An interesting alternative is provided by multiconfigurational quantum chemistry methods, which are particularly well suited for accurately describing the partially filled and strongly localized 4$f$ states\cite{Dubrovin2021,Shaikh2024,Karimi2025}

Strong magnetic anisotropy (MA) is a necessary condition for a stable adatom magnetic moment. In the case of lanthanide adatoms, their strong spin–orbit coupling (SOC) often results in a large MA barrier, especially when the adatom resides at a high-symmetry site~\cite{Donati2016,Baltic2016,Singha2021}. For transition metal adatoms, although number of systems with substantial MA have been reported (e.g.,~\cite{Gambardella2003,Baumann2015}), realizing significant MA is generally more challenging due to their much weaker SOC.

For practical applications, the ability to exert additional control over adatom properties, particularly their structural, electronic, and magnetic characteristics, via an applied electric field is essential for advancing data storage and quantum logic devices. Notably, coherent manipulation of a single nuclear spin using an electric field has been demonstrated in the TbPc$_2$ molecule~\cite{Thiele2014}, enabling the implementation of the Grover algorithm~\cite{Godfrin2017}. Since experimentally achievable electric fields are often insufficient to effectively modulate adatom properties, using ferroelectric materials as substrates offers a promising alternative. These materials exhibit intrinsic electric polarization that can be switched by an external electric field, producing strong internal fields and local structural changes that modify the adatom–substrate bonding and influence the adatom properties. In fact, for thin magnetic layers deposited on ferroelectric substrates, efficient control of magnetization and MA through reversible polarization switching has already been demonstrated (see Ref.~\cite{Birol2012} for a review). Extending this approach to individual magnetic adatoms represents a promising direction for future research.

\begin{figure}
\centering
\includegraphics[width=0.5\linewidth]{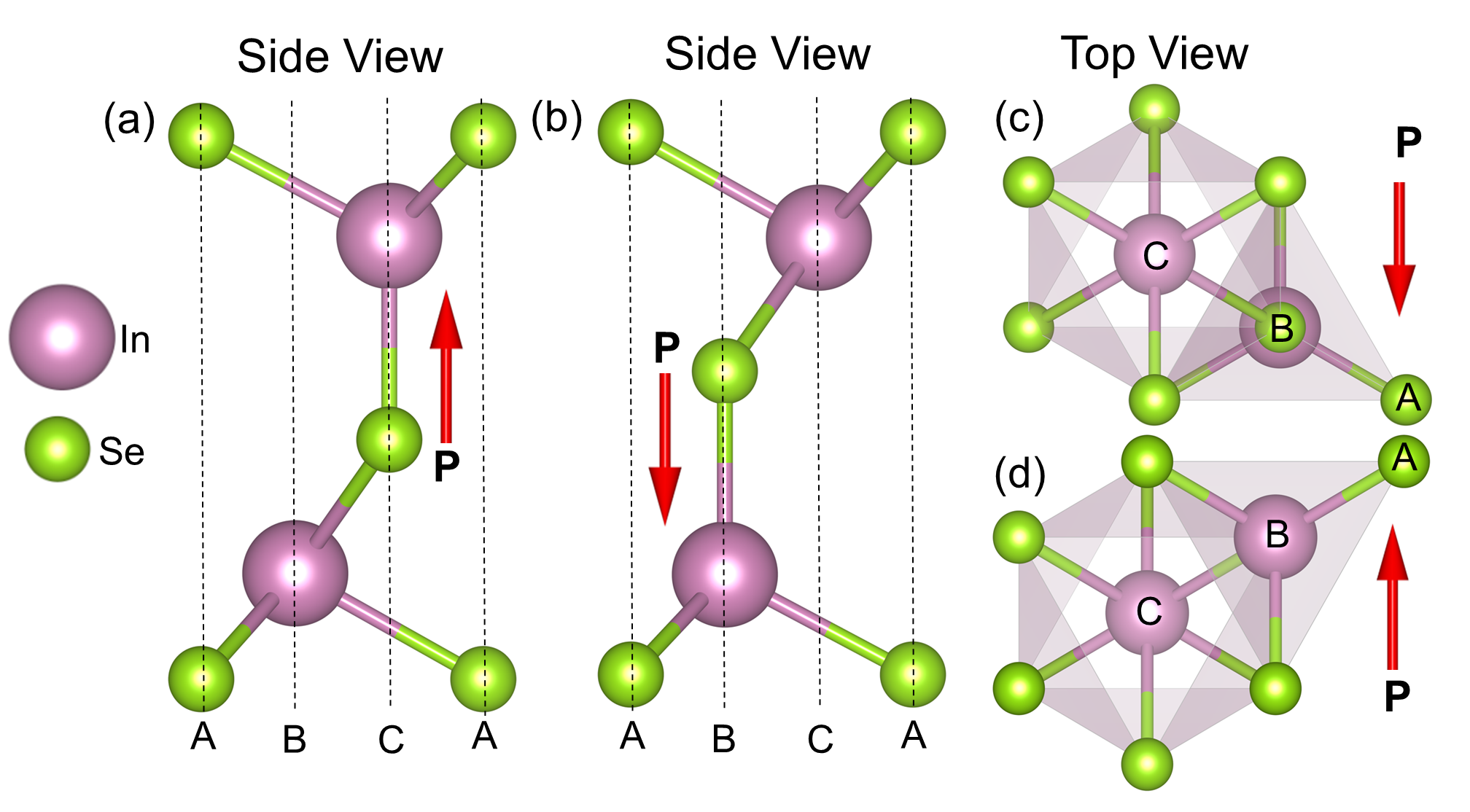}\vspace{-0pt}
\caption {Crystal structure of the ferroelectric $\alpha$-In$_2$Se$_3$ monolayer. Panels (a) and (b) show side views with the electric polarization pointing upward and downward, respectively. Panels (c) and (d) show the corresponding top views. Atomic sites A, B, and C are indicated.} 
\label{In2Se3}
\end{figure}

$\alpha$-In$_2$Se$_3$ is a two-dimensional ferroelectric material that exhibits out-of-plane electric polarization~\cite{Ding2017,Zhou2017,Xue2018,Huang2022}. The crystal structure of a monolayer with both polarization orientations is shown in Figure~\ref{In2Se3}. As an indirect band gap semiconductor~\cite{Liu2021}, $\alpha$-In$_2$Se$_3$ lacks a high density of conduction electrons, which are known to facilitate spin relaxation in magnetic adatoms~\cite{Donati2021}. This makes $\alpha$-In$_2$Se$_3$ a promising substrate for hosting magnetic adatoms\cite{Jin2023}.

In this work, we investigate the potential of using ferroelectric materials as substrates for magnetic adatoms by studying Fe adatoms on $\alpha$-In$_2$Se$_3$. Using first-principles electronic structure calculations, we examine the structural, electronic, and magnetic properties of this system and their dependence on the orientation of the substrate's electric polarization. We identify favorable adsorption sites and analyze the configurational dynamics of the adatoms. The electronic structure, magnetic moments, and MA of Fe adatoms are then explored for different adatom concentrations, with particular attention to how these properties are influenced by the polarization direction of the $\alpha$-In$_2$Se$_3$ substrate.

\begin{figure*}
\centering
\includegraphics[width=\linewidth]{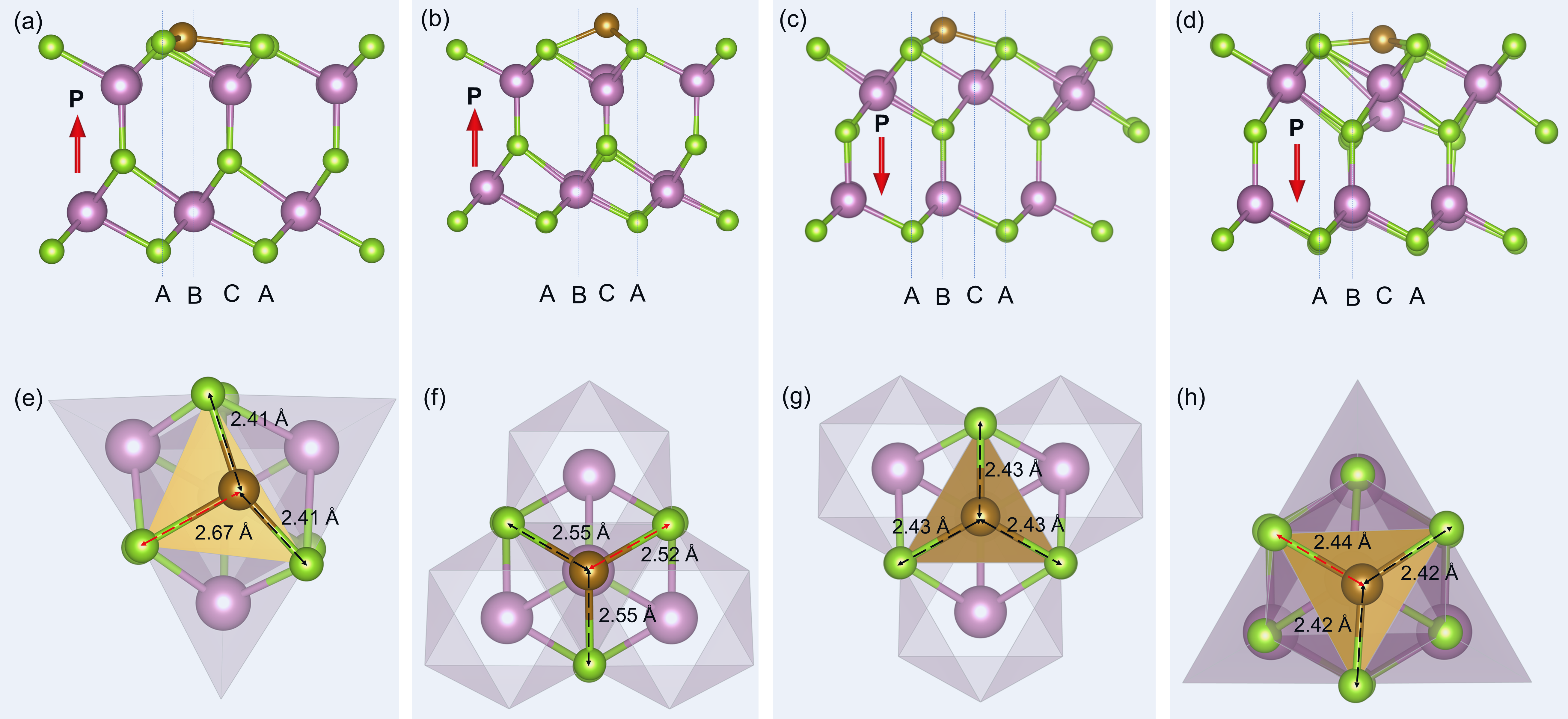}\vspace{-0pt}
\caption {Preferred Fe adatom adsorption sites on the $\alpha$-In$_2$Se$_3$ monolayer. Panels (a) and (e) show the side and top views, respectively, of the Fe adatom at the B adsorption site with the electric polarization pointing up. Panels (b) and (f) show the same views for the C site with polarization up. Panels (c) and (g) show the B site with polarization down, and panels (d) and (h) show the C site with polarization down. Brown, green, and pink spheres represent Fe, Se, and In atoms, respectively. The structures correspond to relaxed configurations in a $6\times6$ lateral supercell. The bottom panels display the relaxed bond lengths between the Fe adatom and its nearest Se neighbors. A Jahn–Teller distortion breaks the threefold symmetry of the adsorption site in all cases except for the B site with polarization down.} 
\label{Feadatom}
\end{figure*}

\section{RESULTS AND DISCUSSION}

{\bf \large $\alpha$-In$_2$Se$_3$ Monolayer}

$\alpha$-In$_2$Se$_3$ has a trigonal, non-centrosymmetric $P3m1$ space group. It is a ferroelectric material with electric polarization along the three-fold axis (out-of-plane direction). The crystal structure of $\alpha$-In$_2$Se$_3$ monolayer is shown in Fig.~\ref{In2Se3}. It consists of five atomic layers arranged in the sequence Se–In–Se–In–Se. Atoms within each layer occupy one of three triangular lattice sites labeled A, B, and C in Fig.\ref{In2Se3}. Specifically, the Se atoms in the outermost layers occupy the A sites, while the In atoms in the second and fourth layers (counting from the bottom) occupy the B and C sites, respectively. The Se atom in the middle layer occupies the C site in the polarization-up state and the B site in the polarization-down state.

We calculated the structural parameters of a free-standing monolayer of In$_2$Se$_3$. The equilibrium in-plane lattice constants are $a = b = 4.10$~\text{\AA}, and the layer thickness along the crystallographic $c$-axis is 6.80~\text{\AA}. These values are consistent with previously reported results~\cite{Ding2017,Liu2021}. The calculated electronic structure of the $\alpha$-In$_2$Se$_3$ mononolayer is shown in Fig.~S1 in Supporing Information. The calculated band gap is approximately 0.78~eV, in agreement with prior PBE calculations~\cite{Liu2021}. The top of the valence band is primarily composed of Se 4$p$ states, with significant contributions from In 5$p$ states. The bottom of the conduction band consists of both In 5$s$ and Se 4$p$ states.
\\

{\bf \large Adsorption Sites}

\begin{figure}
\centering
\includegraphics[width=0.5\linewidth]{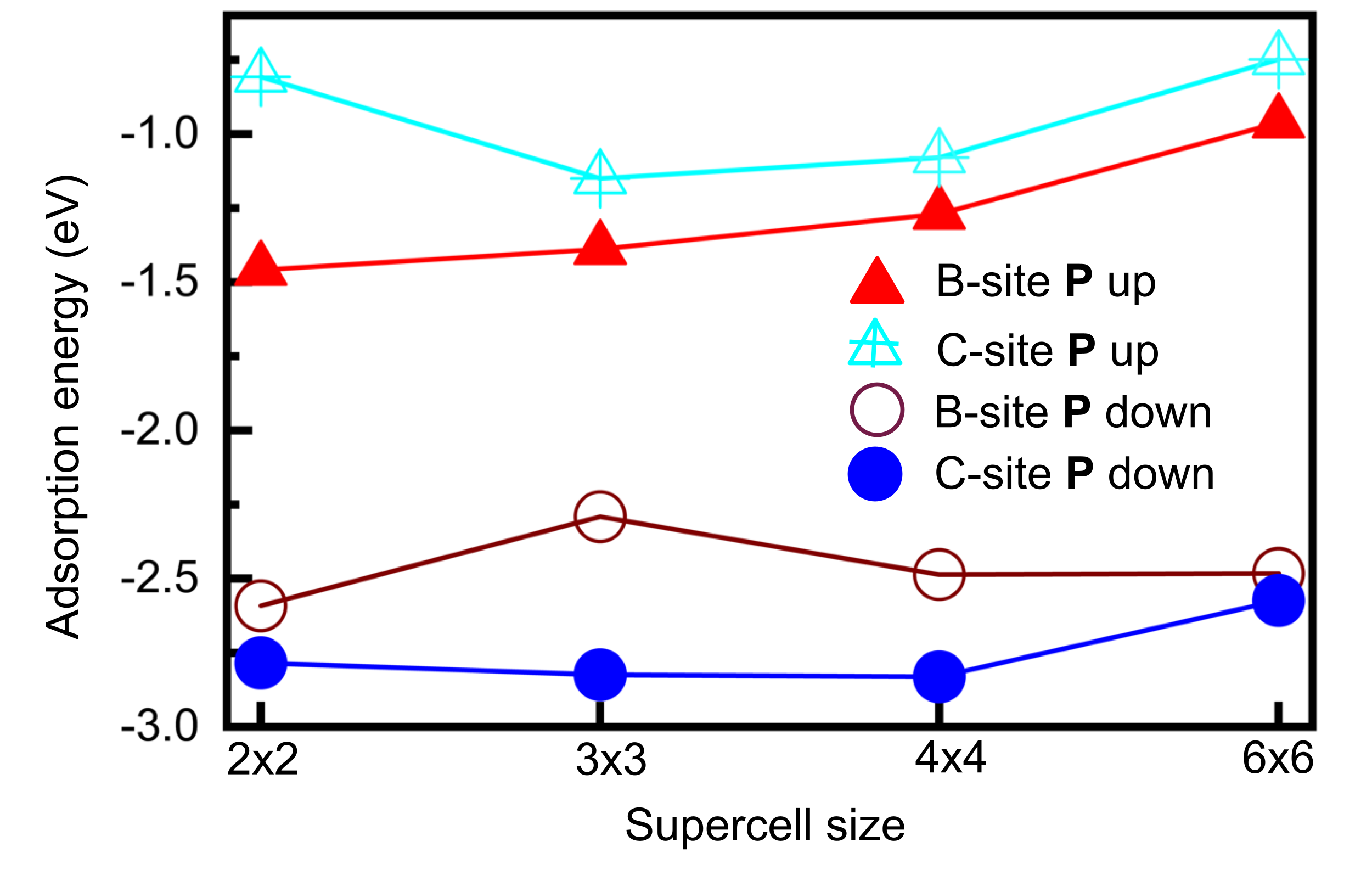}\vspace{-0pt}
\caption {Adsorption energy of the adatom at the B and C sites as a function of lateral supercell size (i.e., adatom concentration), for both directions of the In$_2$Se$_3$ electric polarization.}
\label{AdsorptionEnergy}
\end{figure}

The A, B, and C sites located above the top Se atomic layer are considered as possible adsorption sites for Fe adatoms. The adsorption energy at each site is calculated using the expression:
\begin{equation}
E_a = E(\text{Fe}|\text{In}_2\text{Se}_3) - M \times E(\text{In}_2\text{Se}_3) - E(\text{Fe}),
\end{equation}
where $E(\text{Fe}|\text{In}_2\text{Se}_3)$ is the total energy of the supercell with the Fe adatom adsorbed, $E(\text{In}_2\text{Se}_3)$ is the energy per formula unit of the pristine In$_2$Se$_3$ monolayer, and $E(\text{Fe})$ is the energy of an isolated Fe atom. The parameter $M$ represents the number of In$_2$Se$_3$ formula units in the supercell. Our results indicate that the B and C sites (see Fig.~\ref{Feadatom}) are the most energetically favorable for Fe adsorption, while the A site has a significantly higher adsorption energy (see Table~T1 in Supporting Information). Figure~\ref{AdsorptionEnergy} shows the adsorption energies of Fe adatoms at the B and C sites as a function of lateral supercell size (i.e., adatom concentration), for both directions of the electric polarization of the In$_2$Se$_3$ monolayer. Overall, adsorption energies are consistently lower when the polarization points downward. This trend arises because, under downward polarization, there is no Se atom located directly beneath the In atoms in the second-from-top atomic layer (see Fig.~\ref{In2Se3}). As a result, these nearby In atoms can relax downward more easily, allowing the adatom to be accommodated more efficiently.

The adsorption energies at the B and C sites are generally comparable, with their relative stability depending on the polarization direction. For upward polarization, the B site is more favorable, whereas for downward polarization, the C site consistently yields the lowest adsorption energy across all considered concentrations.

These findings suggest that the direction of electric polarization in In$_2$Se$_3$ can be used to selectively control adatom site preference during Fe deposition. Specifically, at high adatom concentrations, upward polarization promotes adsorption at the B site, while downward polarization favors the C site. This implies that controlling the substrate polarization via an external electric field offers a pathway for directing Fe adatom placement during growth.
\\

{\bf \large Atomic Relaxation}

Table T2 in Supporting Information shows optimized vertical distances of the Fe adatom above the In$_2$Se$_3$ surface for both adsorption sites (B and C) and polarization directions as a function of the lateral supercell size. Let's first consider the $6\times6$ lateral supercell~—~the lowest concentration case considered, which approximates the single-adatom limit. For the polarization-up case, the Fe adatom at the B site is positioned very close to the surface, just 0.16~\AA\ above the top Se layer. In contrast, at the C site, the adatom sits significantly higher~—~nearly 1.0~\AA\ above the surface. This behavior reflects the underlying atomic structure: there are no atoms directly beneath the B site, while an In atom lies directly below the C site (see Fig.~\ref{Feadatom}). In the polarization-down configuration, the trend is reversed. Despite the absence of atoms beneath the B site, the adatom resides relatively high above the surface (0.65~\AA). This elevated position is likely due to the absence of a Jahn-Teller distortion in this configuration (as discussed below), resulting in weaker bonding with surface Se atoms. Meanwhile, at the C site, the adatom sits much closer to the surface (0.20~\AA), even though an In atom lies directly beneath it. This is possible because, under downward polarization, the underlying In atom lacks a supporting Se atom underneath (see Fig.~\ref{Feadatom}), allowing it to relax downward and create space for the Fe adatom. As the adatom concentration increases, this relaxation behavior remains qualitatively unchanged, with only a slight increase in the adatom–surface distance.

Both the B and C adsorption sites on the $\alpha$-In$_2$Se$_3$ monolayer have threefold rotational symmetry about an axis perpendicular to the layer plane. However, adsorption of an Fe atom at these sites can break this symmetry. In fact, we find that~—~except for the B site under downward polarization~—~the Fe adatom slightly shifts away from the high-symmetry position. This displacement results in unequal bond lengths between the Fe atom and its three nearest-neighbor Se atoms (see the bottom panel of Fig.~\ref{Feadatom}). This symmetry breaking arises from the Jahn-Teller effect. As discussed in more detail below, the $3d$ electronic configuration of the Fe adatom typically leads to an orbitally degenerate ground state when placed at a high-symmetry, threefold coordinated site. To lower its energy, the system undergoes a Jahn-Teller distortion, shifting the Fe atom away from the symmetric position. The effect is most pronounced for the B site under upward polarization in the $6\times6$ lateral supercell, as shown in Fig.~\ref{Feadatom}. In contrast, the distortion is much weaker at the C site for both polarization directions. Notably, no Jahn-Teller distortion is observed for the B site under downward polarization, and the Fe atom remains located at the threefold symmetric position.
\\

{\bf \large Adatom Hopping Energy Barriers}

\begin{figure*}
\centering
\includegraphics[width=\linewidth]{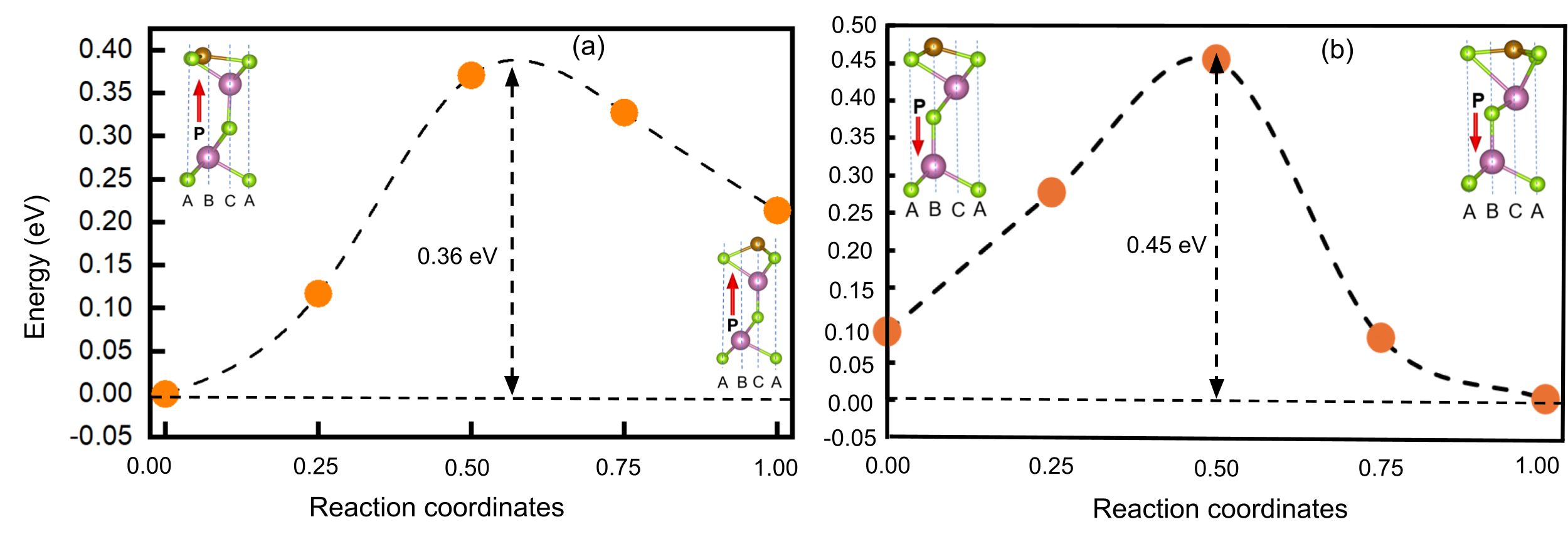}\vspace{-0pt}
\caption {Energy barriers for thermally activated hopping of the Fe adatom between B and C adsorption sites, calculated using the CI-NEB method for the $6 \times 6$ lateral supercell. Results are shown for In$_2$Se$_3$ with electric polarization (a) upward and (b) downward.}
\label{Hopping}
\end{figure*}

The presence of two competing adsorption sites with similar energies may lead to thermally activated hopping of Fe adatoms between them. To assess the significance of this process, we compute the energy barriers for adatom hopping between the B and C sites using the CI-NEB method~\cite{henkelman2000climbing}. The calculations are performed using the 6$\times$6 lateral supercell (corresponding to low adatom concentration), with three intermediate images inserted between the energy minima at the B and C sites. Smooth energy profiles are obtained for both polarization directions, as shown in Fig.~\ref{Hopping}. The computed energy barriers are 0.36~eV and 0.45~eV for the polarization-up and polarization-down states, respectively. These values represent the activation energy $E_b$ for adatom hopping.

The hopping frequency can be estimated using the Arrhenius equation, $\gamma = \gamma_0 e^{-E_b / k_BT}$, where $\gamma_0$ is the attempt frequency, typically on the order of a phonon frequency ($\gamma_0 \sim 10^{13}$~s$^{-1}$). At room temperature, the estimated hopping times are approximately 0.1~ms and 1~ms for the polarization-up and -down states, respectively. As the temperature decreases, the hopping time increases exponentially, and below a characteristic blocking temperature, the hopping becomes effectively frozen. The calculated activation energies correspond to a blocking temperature of approximately 100~K.

\begin{figure*}[t]
\centering
\includegraphics[width=\linewidth]{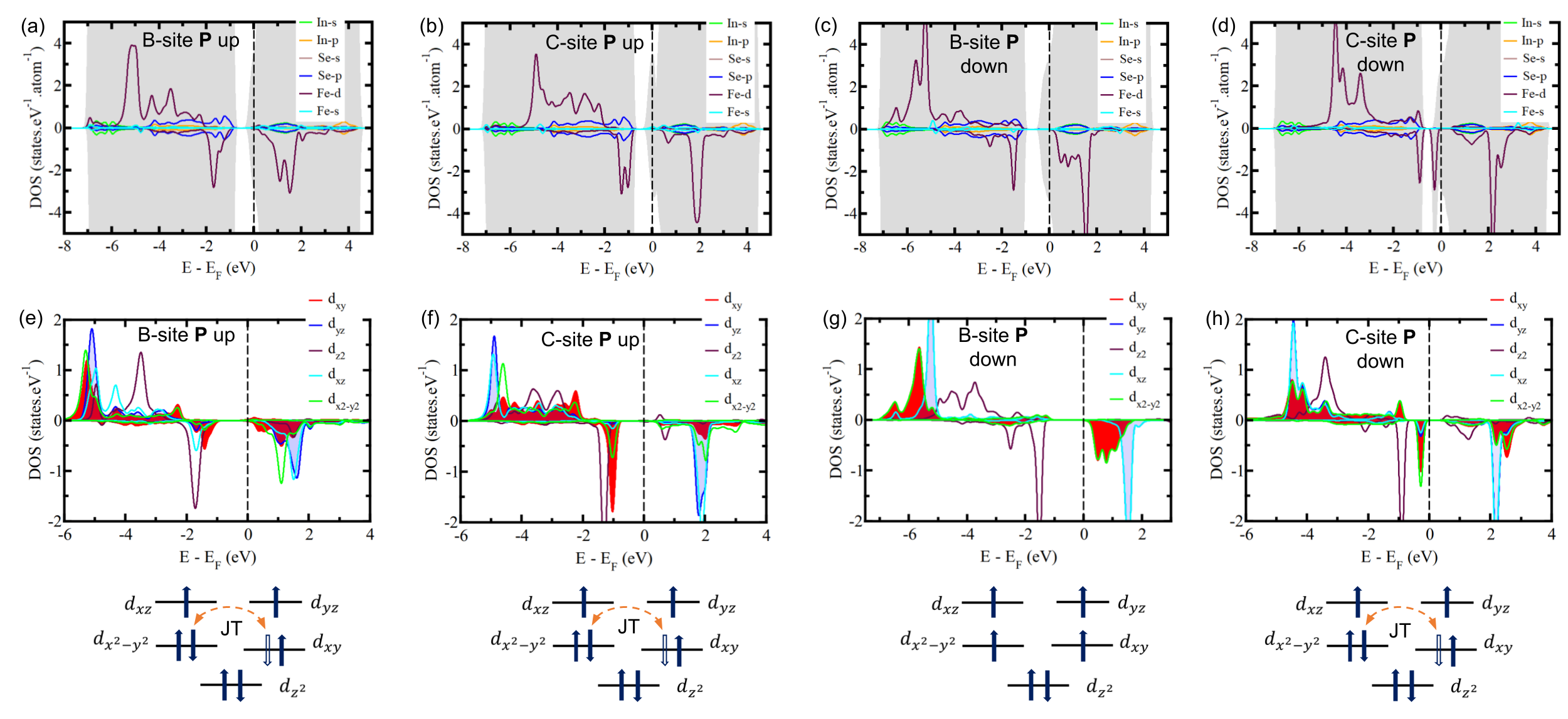}\vspace{-0pt}
\caption{Partial density of states (DOS) of an Fe adatom on In$_2$Se$_3$, calculated using a $6 \times 6$ lateral supercell for B and C adsorption sites and both polarization directions. In the top row, the gray-shaded regions represent the total DOS (scaled down by a factor of 36). The bottom row shows Fe $3d$ single-particle levels split by the ideal (i.e., no Jahn–Teller distortion) threefold-symmetric trigonal crystal field, along with their electronic occupations. Orbital degeneracy responsible for the Jahn–Teller distortion is indicated. Note that the level ordering in the middle row differs from the single-particle diagram due to the presence of the Hubbard $U$ interaction (see text).}
\label{DOS}
\end{figure*}

These results indicate that below 100~K, or under cryogenic conditions, the thermal motion of Fe adatoms is effectively frozen. As a result, the substrate polarization can be switched without inducing hopping between adsorption sites. This stability enables the use of polarization as a tuning parameter to modulate the electronic and magnetic properties of the adatoms without altering their positions. We explore these polarization-driven effects in the following sections.
\\

{\bf \large Electronic Structure}

Figure~\ref{DOS} displays the electronic structure of Fe adatoms adsorbed at the B and C sites on the In$_2$Se$_3$ monolayer, computed using a $6 \times 6$ lateral supercell for both polarization directions. The top row (panels a–d) shows the total density of states (DOS) and the projected DOS resolved by atomic species and orbital character. The middle row (panels e–h) presents the projection of the Fe 3$d$ orbitals onto cubic harmonics. The bottom row illustrates the crystal field-split Fe 3$d$ single-particle levels (not including the Hubbard $U$ interaction) under an ideal threefold-symmetric trigonal crystal field, along with their corresponding electronic occupations.

At such low adatom concentrations, interactions between adatoms are minimal, placing the system close to the single-adatom limit. We first focus on the Fe 3$d$ single-particle levels, which are split by the crystal field generated by the In$_2$Se$_3$ substrate. For Fe atoms adsorbed at the three-fold symmetric B or C sites (neglecting any Jahn-Teller distortion), the local crystal field has a trigonal symmetry. This symmetry splits the 3$d$ orbitals into a singlet ($d_{z^2}$), a doublet ($d_{x^2-y^2}$, $d_{xy}$), and another doublet ($d_{xz}$, $d_{yz}$), with the $z$-axis oriented perpendicular to the In$_2$Se$_3$ plane. The $d_{z^2}$ singlet lies lowest in energy, primarily due to its strong hybridization with the nominally empty In 5$s$ and 5$p$ orbitals near the adatom. This interaction lowers its energy relative to the other 3$d$ states. In contrast, the ($d_{x^2-y^2}$, $d_{xy}$) and ($d_{xz}$, $d_{yz}$) doublets interact with the  nominally occupied Se 4$p$ orbitals at the surface. These interactions shift both doublets to higher energies, with the ($d_{xz}$, $d_{yz}$) levels slightly higher due to stronger hybridization resulting from the Fe adatom’s position above the Se surface layer. The resulting energy ordering of the Fe 3$d$ single-particle levels is illustrated schematically at the bottom of Fig.\ref{DOS}. Note that due to hybridization with substrate orbitals, these discrete levels broaden into bands.

An isolated Fe atom has two 4$s$ and six 3$d$ valence electrons. When adsorbed onto the In$_2$Se$_3$ surface, the six 3$d$ electrons occupy the crystal-field-split 3$d$ levels according to the level diagram shown at the bottom of Fig.~\ref{DOS}. Due to strong intra-atomic exchange interaction (Hund's first rule), the five majority-spin levels are filled first. The remaining electron occupies the minority-spin $d{z^2}$ orbital. As a result of the on-site Coulomb repulsion (Hubbard $U$), this doubly occupied $d{z^2}$ orbital is shifted upward in energy but retains its exchange splitting (see the middle row of Fig.~\ref{DOS}). In solid-state environments, Fe 4$s$ electrons are often ionized, either transferring to the Fe 3$d$ shell or to neighboring atoms. A similar behavior is observed for Fe adatoms on In$_2$Se$_3$. For the B site under downward polarization, the Fe atom retains a $3d^6$ configuration, and most of its 4$s$ electrons are donated to the In$_2$Se$_3$ conduction band, which is primarily composed of Se 4$p$ and In 5$s$ states. Consequently, the surface becomes metallic with the Fermi level lying well within the conduction band. In contrast, for the B site under upward polarization and for the C site under both polarization directions, one of the 4$s$ electrons is transferred to the Fe 3$d$ shell, resulting in a $3d^7$ configuration. In these cases, less charge is donated to the substrate, and although the Fermi level still resides in the conduction band, it is lower than in the B-site downward polarization case. 

The additional electron in the $3d^7$ configuration occupies one of the minority-spin orbitals of the orbitally degenerate ($d_{x^2-y^2}$, $d_{xy}$) doublet, as indicated by the Fe 3$d$ level diagram. This degeneracy makes the system susceptible to a Jahn-Teller distortion, causing the Fe adatom to shift away from the ideal symmetric adsorption site. This distortion breaks the local trigonal symmetry, lifting the degeneracy of the ($d_{x^2-y^2}$, $d_{xy}$) pair. The extra electron then occupies the lower-energy linear combination of these two orbitals, while the higher-energy combination remains unoccupied. The increased occupancy of the ($d_{x^2-y^2}$, $d_{xy}$) states leads to an upward energy shift of this doublet due to the Hubbard $U$ interaction (see the middle row of Fig.~\ref{DOS}). As discussed above, the Jahn–Teller distortion is strongest for the B site in the polarization-up case, leading to significant mixing of the $3d$ orbitals due to large deviations from trigonal symmetry.

While the qualitative ordering of the 3$d$ levels remains consistent across all adsorption sites and polarization directions, the extent of hybridization—and consequently the bandwidth and level positions—varies depending on both the adsorption site and the direction of electric polarization. In particular, for the C site with the polarization down, the Hubbard $U$ interaction pushes the occupied linear combination of the ($d_{x^2-y^2}$, $d_{xy}$) orbitals into the In$_2$Se$_3$ band gap, just below the Fermi level, effectively closing the gap.

\begin{figure*} [t]
\centering
\includegraphics[width=\linewidth]{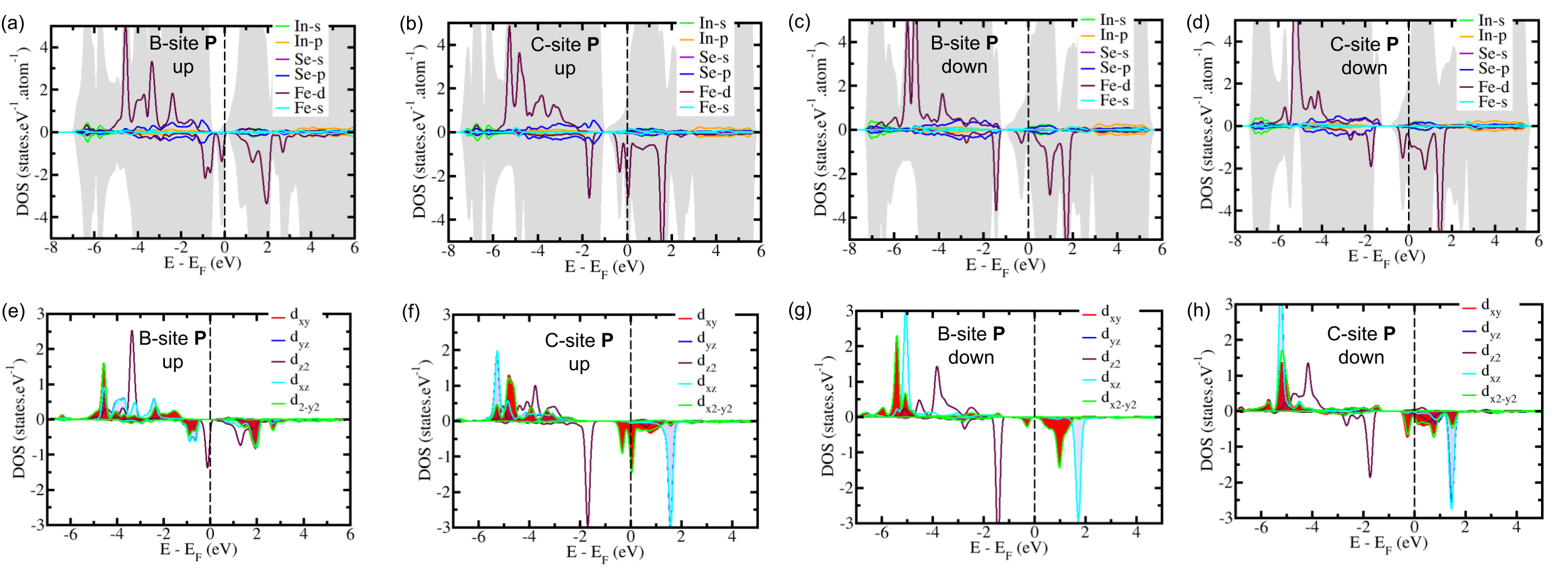}\vspace{-0pt}
\caption{Partial density of states (DOS) of an Fe adatom on In$_2$Se$_3$, calculated using a $2\times2$ lateral supercell for B and C adsorption sites and both polarization directions. The grey shaded regions in the top row denote the total DOS (scaled down by a factor of 4).}
\label{22DOS}
\end{figure*}

As the concentration of Fe adatoms increases, the qualitative features of the electronic structure initially remain unchanged. Specifically, for the $4\times4$ and $3\times3$ lateral supercells (see DOS in Figs.~S2 and S3 in the Supporting Information), Fe adatoms at the C site (both polarization directions) and at the B site with polarization up exhibit a $3d^7$ configuration with a Jahn–Teller distortion, whereas the B site with polarization down yields a $3d^6$ configuration and preserved trigonal symmetry. For even higher adatom concentrations, corresponding to the $2\times2$ lateral supercell, the electronic structure can change significantly as shown in Fig.~\ref{22DOS}. While for the B site, the qualitative features remain consistent with those found at lower concentrations, DOS for the C site differs markedly from that observed at lower adatom concentrations. Indeed, for the Fe adatom at the C site (for both polarization directions), the ($d_{x^2-y^2}$, $d_{xy}$) orbitals form a partially occupied band, with the Fermi level lying inside it, signaling a significantly more delocalized and metallic character of the adatom $3d$ states compared to the more localized states at lower adatom concentrations.

\begin{table}[t]
\setlength{\tabcolsep}{3 pt} 
\begin{center}
\caption{\label{Magnetism} Magnetic properties as a function of adatom concentration, adsorption site, and polarization direction. Here, $M_S$ is the spin magnetic moment, $M_L^z$ ($M_L^x$) is the orbital magnetic moment with the spin quantization axis oriented out-of-plane (in-plane), and MAE is the magnetic anisotropy energy.}
\begin{tabular}{c|l|cccc}
\hline
\hline
Supercell         & Adsorption site        & $M_S$ ($\mu_B$) & $M_L^z$ ($\mu_B$) & $M_L^x$ ($\mu_B$) & MAE (meV) \\
\hline
$6\times6\times1$ & B-site \textbf{P} up   & 3.03            & 0.10               & 0.15             & -0.66     \\
                  & C-site \textbf{P} up   & 2.97            & 0.17               & 0.13             &  0.55     \\
                  & B-site \textbf{P} down & 3.48            & 0.06               & 0.14             & -0.77     \\
                  & C-site \textbf{P} down & 3.07            & 0.17               & 0.10             &  1.23     \\
\hline
$4\times4\times1$ & B-site \textbf{P} up   & 3.03            & 0.10               & 0.14             & -0.60     \\
                  & C-site \textbf{P} up   & 2.98            & 0.16               & 0.13             &  0.38     \\
                  & B-site \textbf{P} down & 3.47            & 0.08               & 0.14             & -0.20     \\
                  & C-site \textbf{P} down & 3.07            & 0.17               & 0.10             &  1.15     \\
\hline
$3\times3\times1$ & B-site \textbf{P} up   & 3.16            & 0.09               & 0.14             & -0.30     \\
                  & C-site \textbf{P} up   & 2.99            & 0.16               & 0.13             &  0.36     \\
                  & B-site \textbf{P} down & 3.44            & 0.17               & 0.14             &  2.62     \\
                  & C-site \textbf{P} down & 3.08            & 0.16               & 0.10             &  1.01     \\
\hline
$2\times2\times1$ & B-site \textbf{P} up   & 2.73            & 0.11               & 0.10             & 0.63      \\
                  & C-site \textbf{P} up   & 3.15            & \textbf{1.21}      & \textbf{0.15}    & \textbf{213.67} \\
                  & B-site \textbf{P} down & 3.39            & 0.11               & 0.14             &  0.61     \\
                  & C-site \textbf{P} down & 3.13            & 0.18               & 0.10             &  1.59     \\
\hline
\hline
\end{tabular}
\end{center}
\end{table}

The Fe adatom magnetic moment as a function of concentration, adsorption site, and polarization direction is shown in Table~\ref{Magnetism}. For the C site, the magnetic moment is close to 3~$\mu_B$ across all considered adatom concentrations and both polarization orientations, consistent with a $3d^7$ electronic configuration. Note that for the high adatom concentration case ($2\times2$ lateral supercell) the deviations from the 3~$\mu_B$ value become more pronounced due to partially metallic nature of the adatom $3d$ states. In contrast, for the B site, both the $3d$ occupancy and resulting magnetic moment exhibit a strong dependence on the polarization direction. When the polarization points downward, the Fe adatom adopts a $3d^6$ configuration, which nominally corresponds to a magnetic moment of 4~$\mu_B$. However, the calculated value is reduced to approximately 3.5~$\mu_B$, as part of the spin density from the more delocalized Fe $3d$ states lies outside the atomic sphere used for the magnetic moment integration. For the polarization pointing upward, the $3d^7$ configuration is restored, and the magnetic moment is around 3~$\mu_B$. These results demonstrate that the electronic and magnetic properties of the Fe adatom can be effectively tuned by switching the substrate's electric polarization.
\\

{\bf \large Magnetic Anisotropy}

For applications of magnetic adatoms in magnetic memory and quantum information technologies, the stability of the adatom magnetic moment against thermal and quantum fluctuations is a crucial factor. Magnetic anisotropy energy (MAE) is defined as the energy difference between the configuration with the magnetic moment aligned in-plane and the one with the magnetic moment oriented out-of-plane.  This quantity determines the barrier height for the magnetic moment reversal and  therefore serves as a key figure of merit for assessing the stability of the magnetic moment.

The dependence of MAE and orbital magnetic moments for different orientations of the spin quantization axis on adatom concentration, adsorption site, and polarization direction is summarized in Table~\ref{Magnetism}. In most cases, we find a sizable MAE of the order of 1 meV. However, both the magnitude and even the sign of MAE vary depending on the adsorption site, concentration, and polarization orientation. In general, the sign of MAE follows the sign of the orbital moment anisotropy, defined as $\Delta M_L^z = M_L^z - M_L^x$. Interestingly, our results also indicate that the sign of MAE is not solely determined by the adatom $3d$ configuration.

For the $6\times6$ lateral supercell, the B site exhibits a negative MAE (favoring in-plane alignment of the magnetic moment, parallel to the In$_2$Se$_3$ surface), whereas the C site shows a positive MAE (favoring out-of-plane alignment). The magnitude of MAE is further modulated by the electric polarization direction. A similar trend is observed for the $4\times4$ lateral supercell. At higher adatom concentrations, however, the behavior changes significantly. In the $3\times3$ lateral supercell, the MAE at the B site depends strongly on polarization: for polarization up it is large and positive (~2.6 meV), while for polarization down it becomes small and negative (~–0.3 meV). Thus, the stability of the adatom magnetic moment can be effectively tuned by switching the polarization direction.

Even more intriguing behavior appears for the $2\times2$ lateral supercell. In this high adatom concentration case, we obtain a gigantic positive MAE (~214 meV) for the C site with polarization up. Such an unusually large value for a $3d$ system originates from the emergence of a sizable orbital magnetic moment (~1.2 $\mu_B$) when the spin quantization axis is oriented out-of-plane. This moment gives rise to a large first-order contribution to the SOC energy, $\sim\lambda SL$, where $\lambda$ is the SOC constant (50–80 meV for $3d$ atoms), while $S \approx 3/2$ and $L \approx 1$ are the spin and orbital angular momentum quantum numbers of the Fe adatom, respectively. In contrast, for the in-plane spin quantization axis, the orbital magnetic moment is nearly quenched, making the first-order SOC contribution negligible and resulting in the colossal MAE. For the C site with polarization down, however, the orbital moment remains small for both orientations of the spin quantization axis, yielding only a moderate MAE (~1.6 meV). Thus, the very strong out-of-plane anisotropy of Fe adatoms can be effectively switched on and off by reversing the electric polarization direction.

\section{CONCLUSIONS}

In summary, we have investigated Fe adatoms on ferroelectric $\alpha$-In$_2$Se$_3$ with out-of-plane electric polarization using first-principles calculations. Our results reveal two competing adsorption sites, and which site has the lower adsorption energy varies with the adatom concentration. Most importantly, we find that the preferred adsorption site can be switched by reversing the electric polarization of the $\alpha$-In$_2$Se$_3$ substrate. This suggests that the specific adsorption site can be selected by fixing the substrate polarization direction during the adsorption process. CI-NEB calculations show that the energy barrier for thermally activated hopping between these adsorption sites is approximately 0.4~eV, corresponding to a blocking temperature of around 100~K, indicating that the selected site can remain stable under cryogenic conditions.

The hybridization between Fe and In$_2$Se$_3$ orbitals depends strongly on the adsorption site and the polarization direction, due to differences in local adatom geometry. As a result, the adatom electronic configuration, magnetic moment, and magnetic anisotropy vary significantly with adsorption site and can be further modulated by switching the electric polarization. In particular, for higher adatom concentrations, a giant magnetic anisotropy exceeding 200~meV per Fe atom emerges for a specific adsorption site and a polarization orientation, which can be substantially reduced by reversing the polarization. The stability of the adatom magnetic moment against thermal fluctuations can therefore be controlled by switching the electric polarization.

Overall, our results suggest that ferroelectric substrates present a viable approach for voltage-controlled manipulation of magnetic adatom properties via reversible polarization switching. While our study focused on the Fe adatom, the underlying mechanism is not restricted to transition-metal systems. Similar electric-field control can, in principle, be extended to rare-earth adatoms, whose non-4$f$ valence electrons are sensitive to changes in the local atomic structure and electrostatic environment induced by ferroelectric polarization switching. This opens new possibilities for designing multifunctional nanoscale spintronic and quantum devices that exploit the interplay between ferroelectricity and magnetism at the atomic scale.

\section{COMPUTATIONAL METHODS}

The calculations are performed using the density functional theory (DFT) within the simplified (spherically averaged) DFT+U method~\cite{dudarev1998electron} and the PBE exchange-correlation functional~\cite{PBE}. We use an effective Hubbard parameter, $U_{eff}= U-J = 3.0$ eV for the Fe $3d$ shell. Kohn-Sham equations are solved using the projector augmented wave method~\cite{kresse1999ultrasoft} as implemented in Vienna $\textit{ab initio}$ simulation package (VASP)~\cite{vasp}. We use the supercell method with Fe adatom on top of a single two-dimensional In$_2$Se$_3$ monolayer (five atomic layers). Different adatom adsorption sites are considered. The vertical lattice parameter of the supercell is fixed such that the adatom/In$_2$Se$_3$ monolayer structure is separated from its periodic image in the vertical direction by 1.5 nm of vacuum. The lateral lattice parameters of the supercell are fixed to the calculated values for a single (no adataom) $\alpha$-In$_2$Se$_3$ monolayer. Different sizes of lateral unit cells are considered including $2\times2$, $3\times3$, $4\times4$, and $6\times6$. Here, $1\times1$ lateral size corresponds to the primitive lateral supercell for the $\alpha$-In$_2$Se$_3$ crystal structure. Note that there is only one Fe adatom in the supercell, and therefore, different lateral unit cells correspond to different concentrations of Fe adatom on the surface. The plane wave cutoff energy is set to 300 eV and the Brillouin zone integration is performed using $\Gamma$-centered Monkhorst-Pack grids. For the $6\times6$ lateral unit cell, the $2\times2\times1$ $k$-point mesh is used for relaxations and the $6\times6\times1$ $k$-point mesh is used for total energy, density of states, and magnetic anisotropy calculations. For smaller lateral supercells, the $k$-point mesh is scaled accordingly. The atomic positions are relaxed until the Hellmann-Feynman forces are converged to less than 0.01 eV/\(\text{\AA}\). The energy barriers for thermally-activated adatom hopping between adsorption sites are calculated using the climbing image nudged elastic band method (CI-NEB)~\cite{henkelman2000climbing}. Magnetic anisotropy energy (MAE) is calculated by including SOC and calculating total energy (self-consistently) for different orientation of the spin quantization axis with respect to the real space. Crystal structure visualization and structural analysis is done using the VESTA software~\cite{vesta}.

\begin{acknowledgement}
This work is supported by the NSF EPSCoR Cooperative Agreement OIA-2044049, Nebraska's EQUATE collaboration. We acknowledge the University of Nebraska Holland Computing Center for computational resources. 
\end{acknowledgement}




\providecommand{\latin}[1]{#1}
\makeatletter
\providecommand{\doi}
  {\begingroup\let\do\@makeother\dospecials
  \catcode`\{=1 \catcode`\}=2 \doi@aux}
\providecommand{\doi@aux}[1]{\endgroup\texttt{#1}}
\makeatother
\providecommand*\mcitethebibliography{\thebibliography}
\csname @ifundefined\endcsname{endmcitethebibliography}
  {\let\endmcitethebibliography\endthebibliography}{}

\end{document}